\documentclass[traditabstract,letter]{aa} 
\usepackage{graphicx}
\usepackage{natbib}
\usepackage{txfonts}
\begin{document}
   \title{The structured environments of embedded star-forming cores\vspace*{-2mm}}
   
   \subtitle{\thanks{{\it Herschel} is an ESA space observatory, with its
       science instruments PACS, SPIRE, and HIFI provided by
       European-led Principal Investigator consortia, and with
       important participation from NASA.}{PACS and SPIRE mapping of
       the enigmatic outflow source UYSO~1}}

   \author{\hspace*{-0.05cm}H.~Linz\inst{1}
          \and
          O.~Krause\inst{1}
	  \and
          H.~Beuther\inst{1}
	  \and
          Th.~Henning\inst{1}
	  \and
	  R.~Klein\inst{2}
	  \and
	  M.~Nielbock\inst{1}
	  \and
	  B.~Stecklum\inst{3}
	  \and
	  J.~Steinacker\inst{4,1}
	  \and
	  A.~Stutz\inst{1}
          }

   \institute{Max-Planck-Instritut f\"ur Astronomie (MPIA) Heidelberg, 
              K\"onigstuhl 17, D-69117 Heidelberg, Germany\\
              \email{[linz,krause,beuther,henning,nielbock,stein,stutz]@mpia.de}
         \and
             Space Sciences Laboratory, University of California, Berkeley, CA
	     94720, USA,  
	     \email{r\_klein@calmail.berkeley.edu}
	 \and
	      Th\"uringer Landessternwarte Tautenburg (TLS),
	      Sternwarte 5, D-07778 Tautenburg, Germany, \\
	      \email{stecklum@tls-tautenburg.de}
	 \and     
	      LERMA, Observatoire de Paris, 61 Av. de l'Observatoire,
              75 014 Paris, France 
             }

   \date{Received March 31, 2010, accepted May 03, 2010}

 
  \abstract{ The intermediate-mass star-forming core UYSO~1 has 
    previously been
    found to exhibit intriguing features. While deeply embedded and
    previously only identified by means of its (sub-)millimeter emission, it
    drives two powerful, dynamically young, molecular
    outflows. Although the process of star formation has obviously
    started, the chemical composition is still pristine. We present
    {\it Herschel} PACS and SPIRE continuum data of this presumably very
    young region. 
    The now complete coverage of the spectral energy peak allows us to
    precisely constrain the elevated temperature of 26\,--\,28 K for the 
    main bulge of gas associated with UYSO1, which is located at the interface 
    between the hot H {\sc ii} region Sh 2-297 and the cold dark nebula 
    LDN 1657A. Furthermore, the data identify cooler compact far-infrared 
    sources of just a few solar masses, hidden in this neighbouring dark 
    cloud.  }


   \maketitle
%

\section{Introduction}

Star formation occurs predominantly in structured
environments in the case of intermediate and high-mass
star formation. In particular, one often finds different stages of 
evolution located within close proximity. This poses a challenge for 
working on the earliest phases of star formation, which 
can only be revealed with long-wavelength observations.  
To observe these deeply
embedded objects at the spectral peak of their emission, airborne or satellite
missions are necessary, which in the past delivered only a very modest
spatial resolution ($>$20$''$).  We use ESA's new far-IR and sub-millimeter
satellite {\it Herschel} \citep{A&ASpecialIssue-Herschel} --- with its
unprecedented spatial resolution --- to scrutinise a sample of very
young low- and high-mass star-forming cores within our programme
 {``Earliest Phases of Star Formation''} (EPoS, PI: O.~Krause).  
One of these targets is a
core located in the \object{Canis Majoris OB1/R1} region
\citep[e.g.,][]{2009A&A...506..711G} of the outer Galaxy at a distance
of $\sim 1$~kpc. Originally detected as a distinct submillimetre
source in the vicinity of \object{IRAS 07029-1215}
\citep{2004ApJ...602..843F}, it was coyly named ``UYSO~1''
(unidentified young stellar object 1) since it was only detected at
450 and 850 $\mu$m and seemed to have no counterpart at other
wavelengths.  In the optical, the entire region is dominated by the
strong H{\sc ii} region \object{Sh 2-297}, excited by the early B star
\object{HD 53623}. The mid-IR emission is dominated by warm dust
and PAH emission from the associated southern photon-dominated region
(PDR). All these sources, however, do not coincide with UYSO~1,
located more than 1$'$ to the north-west. Further to the west, one
finds the optical dark cloud \object{LDN 1657A}, about which little is
known. Members of our group carried out a 
multiwavelength study of \object{UYSO 1} \citep{2009A&A...493..547F},
which revealed more puzzles.  The source drives a dynamically
young, but strong CO molecular outflow. The imprints of two crossed
jets are visible in shocked H$_2$ emission in the near-IR, and
their vertex is very close to the position of UYSO~1.  Millimeter
continuum interferometry resolved the source into two peaks separated
by 4\farcs2.  {\it Spitzer}/MIPS data indicated that at 24 and 70 $\mu$m, 
UYSO~1 is still too deeply embedded to be directly detected, 
 a finding we revise in the present paper for  
70 $\mu$m with {\it Herschel}.  In addition to the core being
chemically pristine and showing no signs of  more evolved chemistry, 
this object is very interesting for the investigation of the early
evolution of (intermediate-mass) protostars.  

\section{Observations and data reduction}

\begin{table}
\begin{minipage}[t]{\columnwidth}
\caption{Table of positions of relevant objects in the {\it Herschel} field.}\label{Table:new-sources}
\centering
\renewcommand{\footnoterule}{}
\begin{tabular}{lll}
\hline \hline
ID& RA (J2000)    &  Dec (J2000)   \\
~ & [{ h~:~m~:~s}] & [{ $^\circ$~:~$'$~:~$''$}]  \\
\hline
   IRAS 07029-1215     &07:05:16.9   &     -12:20:02   \\
   HD 53623            &07:05:16.75  &     -12:19:34.5 \\
   Sh-2 297            &07:05:15.4   &     -12:19:39   \\
\hline
   UYSO~1$^a$          &07:05:11.14  &   -12:19:00.2 \\
   LDN 1657A-1$^b$     &07:04:57.14  &	 -12:16:57.4 \\
   LDN 1657A-2$^b$     &07:04:58.03  &	 -12:16:50.8 \\
   LDN 1657A-3$^b$     &07:05:00.69  &	 -12:16:44.7 \\
   LDN 1657A-4$^a$     &07:05:03.50  &	 -12:16:33.7 \\
   LDN 1657A-5$^b$     &07:05:05.84  &	 -12:16:51.4 \\
\hline
\end{tabular}
\end{minipage}
$^a $derived from astrometry-adjusted PACS 70~$\mu$m map (see Sect.~\ref{Sect:UYSO1}) \\
$^b $derived from MIPS 24~$\mu$m map (see Sect.~\ref{Sect:new-sources})
\end{table}
\begin{table}
\begin{minipage}[t]{\columnwidth}
\caption{Point source fluxes (without colour correction). The overall 
uncertainties are 20\% for all wavelengths except for 160 $\mu$m (25\%).}\label{Table:fluxes}
\centering
\renewcommand{\footnoterule}{}  
\begin{tabular}{lrrrr}
\hline \hline
ID&  F$_\nu$ [{\rm Jy}] & F$_\nu$ [{\rm Jy}] & F$_\nu$[{\rm Jy}] & F$_\nu$[{\rm Jy}] \\
 &  (24~$\mu$m)    & (70~$\mu$m) &  (100~$\mu$m)   &  (160~$\mu$m)\\
\hline
UYSO~1$^a$      &  --     &	168  &  239  &  193 \\
LDN 1657A-1     &  0.07   &	0.6  &  0.8  &  1.5 \\
LDN 1657A-2     &  0.10   &	0.2  &  0.2  &  $<$0.2 \\
LDN 1657A-3     &  0.09   &	0.5  &  0.9  &  1.3 \\
LDN 1657A-4     & $<$0.002 &	0.3  &  1.1  &  2.6 \\
LDN 1657A-5     &  0.04   &	0.4  &  0.5  &  0.7 \\
\hline
\end{tabular}
\end{minipage}
$^a$ Photometry obtained with large aperture (see Sect.~\ref{Sect:UYSO1}). The corresponding SPIRE
     fluxes used for Fig.~\ref{Fig:UYSO-SED} are:  {(91.1$\pm$13.7)} Jy (250 $\mu$m),  {(32.5$\pm$4.9)} Jy (350 $\mu$m), 
     and  {(11.2$\pm$1.7)} Jy (500 $\mu$m).
\end{table}

   \begin{figure*}
   \centering
   \includegraphics[bb=0 10 600 578,width=7.2cm,clip]{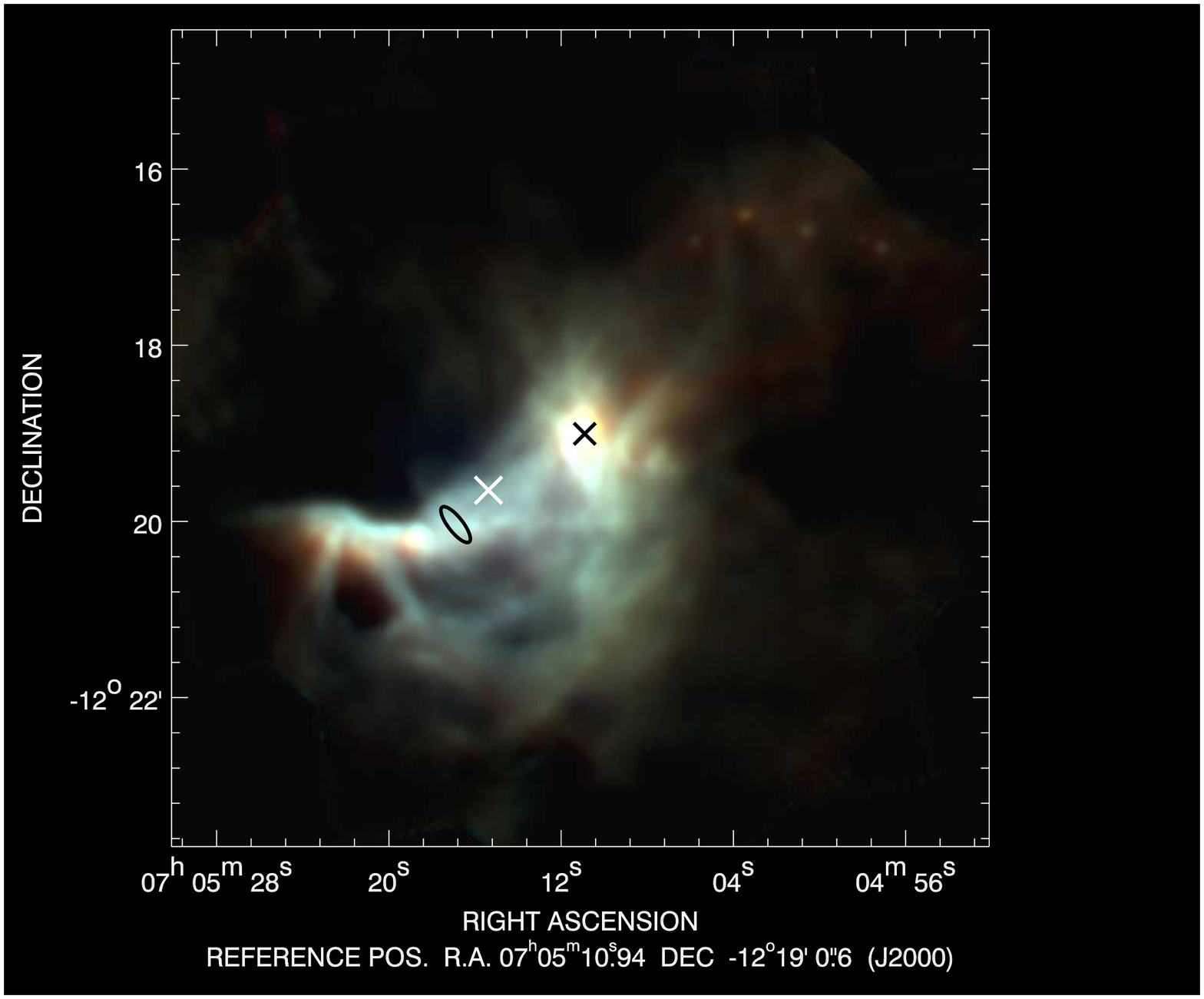}\includegraphics[bb=0 10 600 578,width=7.2cm,clip]{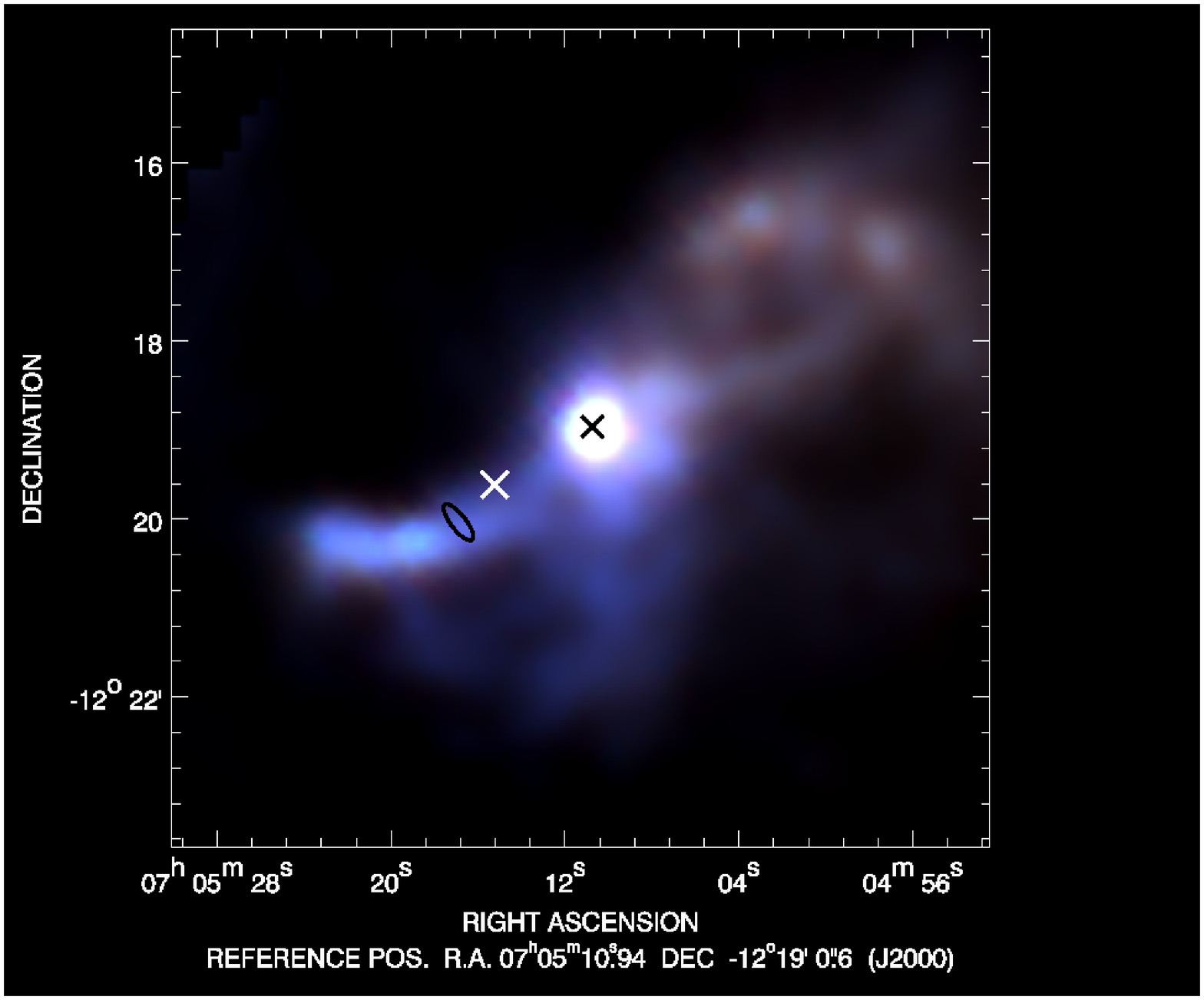}
      \caption{{\bf Left: } Star-forming complex around UYSO~1 with PACS. 
        Three-colour composite with 70 $\mu$m (blue), 100 $\mu$m  (green), 
	and 160 $\mu$m (red). Black cross at 
	reference position: location of UYSO~1 derived by interferometry 
	(see Fig.~\ref{Fig:70micron-centre}), white cross: peak of
	Sh 2-297 cm emission, ellipse: error ellipse of IRAS 07029-1215 
	position.
	{\bf Right: } SPIRE three-colour
        composite with 250 $\mu$m (blue), 350 $\mu$m (green),
        and 500 $\mu$m (red).}
         \label{Fig:PACS-3colour}
   \end{figure*}

The object UYSO~1 and its vicinity were observed in the blue (70
$\mu$m), green (100 $\mu$m), and red (160 $\mu$m) filters with the
PACS instrument \citep{A&ASpecialIssue-PACS} aboard {\it Herschel} on October 9,
2009, within the science demonstration program. Scan maps in two
orthogonal directions with scan leg lengths of 7$\arcmin$ were 
obtained with a scan speed of 20$''$/s,  for which the spatial 
resolution is 5\farcs6, 6\farcs8, and 11\farcs3, respectively. 
The raw data were reduced with the HIPE software
\citep[e.g.,][]{A&ASpecialIssue-PACS}, version 3.0, build 455. Beside
the standard steps leading to level-1 calibrated data, a second-level
deglitching and a correction for offsets in the detector sub-matrices
were performed. Finally, the data were highpass-filtered,
using a median window of the size of the full scan legs, to
remove the effects of bolometer sensitivity drifts during data
acquisition and 1/f noise. We masked out emission structures (visible in 
a first iteration) before computing and subtracting this running median; 
this step minimises oversubtraction of source emission. Finally, the data
were projected onto a coordinate grid using the {\it photProject}
routine inside HIPE, and flux calibration factors \citep{A&ASpecialIssue-PACS}
were applied. \\
\indent
Maps at 250, 350, and 500 $\mu$m were obtained with SPIRE
\citep{A&ASpecialIssue-SPIRE} on October 19, 2009. Two 9$\arcmin$ scan
legs covered the source with the nominal speed of 30$''$/s. The data
were processed within HIPE with the standard photometer script up to
level 1. During baseline removal, we masked out the high-emission area
in the center. For these observations, no cross-scan data was acquired;
therefore the resulting maps still show residual striping along the
scan direction. An iterative destriping algorithm
\citep{A&ASpecialIssue-Bendo} was applied to mitigate this effect.
We adopt FWHM beam sizes of 18\farcs1, 25\farcs2, and 36\farcs9 at 250,
350, and 500 $\mu$m, respectively.

\section{Results}

The {\it Herschel} data show the detailed structure of
the entire star-forming complex containing UYSO1. The data are shown
in Fig.~\ref{Fig:PACS-3colour}, where we have combined the PACS and SPIRE
data into two colour composites.  In the southern region, there is a
ridge of far-IR emission roughly oriented east-west.  The
extended emission there covers the formal position of IRAS 07029-1215.
The IRAS positions, however, have usually been derived from the IRAS
12~$\mu$m maps. We note that the reported position coincides with neither
the much stronger far-IR peak associated with UYSO~1 (see
below) nor the peak position of centimetre emission arising from the
H{\sc ii} region (cf.~Fig.\ref{Fig:PACS-3colour}). 
Parts of this emission ridge represent a
PDR adjoining the H{\sc ii} region.  Nevertheless, the bright condensation
east of the IRAS position within this PDR ridge might correspond to
another embedded compact core. \\
\indent
While this southern ridge is the brightest source at mid-IR
wavelengths ($\lambda\lesssim25$~$\mu$m), the Herschel data present
a far different picture.  The emission area around UYSO~1 dominates
the {\it Herschel} bands: UYSO~1 itself is situated on top of a highly
structured emission plateau \citep[not detected in the SCUBA 
450~$\mu$m data of][]{2004ApJ...602..843F} with a similar extent
and structure as the 850 $\mu$m map by \citet{2004ApJ...602..843F}. 
The emission in the close eastern vicinity ($\sim30''$) of UYSO~1 is also
a PDR excited by the B star as indicated by the [O {\sc i}] line 
observed by \citet{2009A&A...493..547F}. We
do not delve into a detailed study of the temperature distribution
for the whole region \citep[see ][for CB\,244]{A&ASpecialIssue-Stutz}.
However, from inspection of Fig.~\ref{Fig:PACS-3colour} it is apparent
that the extended emission around UYSO1 has its maximum shortwards of
160~$\mu$m, while the region extending to the north-west contains
emission that only becomes pronounced at the SPIRE wavelengths,
indicating that even colder material is located in the LDN 1657A dark
cloud.

\subsection{UYSO~1}\label{Sect:UYSO1}

   \begin{figure}
   \centering
   \includegraphics[bb=30 0 610 558,width=5.cm,clip]{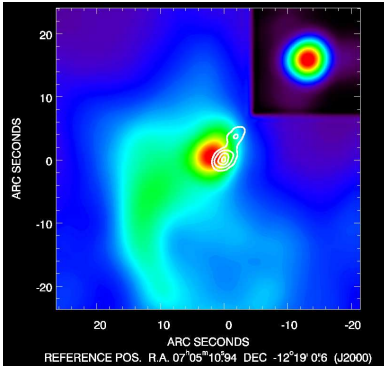}
      \caption{PACS 70 $\mu$m image of the central region of UYSO~1.
        The inset shows the PACS 70 $\mu$m PSF
        \citep{A&ASpecialIssue-PACS}, rotated and scaled to match the
        UYSO~1 observation.  The white contours denote the SMA 349~GHz
        continuum detection (H.~Beuther, priv.~comm.)}
         \label{Fig:70micron-centre}
   \end{figure}
   
At far-IR wavelengths, the UYSO~1 core dominates the continuum
emission of the region. The PACS instrument offers unprecedented
spatial resolution of around 5\farcs6 for the 70 $\mu$m filter
providing the opportunity to relate the far-IR emission to
high-resolution data at other wavelengths.
\citet{2009A&A...493..547F} reported {\it Spitzer}/MIPS observations of
UYSO~1 at 70 $\mu$m as well as 3 mm interferometric data pinpointing
the two continuum peaks in UYSO~1. A large position shift of $> 9\arcsec$ 
was found between the mm interferometric sources and the 70 $\mu$m peak. 
Thus, the authors concluded that the optical depth at the column density 
peak is still too high for a direct detection of the sources at 70 $\mu$m. 
This large discrepancy in peak positions is not confirmed by our PACS
data. Figure~\ref{Fig:70micron-centre} shows the PACS view on the
central region around UYSO~1. As a position reference for  
the two related mm sources, we use new SMA 349 GHz
continuum data (H.~Beuther, priv.~comm.) that are in very good
positional agreement with the 3 mm PdBI data from
\citet{2009A&A...493..547F}.  
The typical 1-$\sigma$ absolute pointing accuracy of
{\it Herschel} is $2\arcsec$; however, deviations of up to 4$''$ have
been reported for a few operational days. 
Fortunately, the PACS data
reveal several point sources northwest of UYSO~1 (not covered by the
MIPS 70~$\mu$m data) that were also detected in the MIPS 24 $\mu$m
map (see Sect.~\ref{Sect:new-sources}). We concatenated the
PACS astrometry to the MIPS 24 $\mu$m astrometry 
(see Fig.~\ref{Fig:70micron-centre}). As a result,
the 70~$\mu$m peak is now close to the stronger of the two
interferometric point sources. The two positions still do not
perfectly coincide, but the remaining offset of $\sim 2\arcsec$ 
is of the order of our final pointing uncertainties.   Thus, 
there is a clear close association between the 70~$\mu$m peak and the
millimetre source(s). As a test, we smoothed our PACS 70 $\mu$m
data to the MIPS 70 $\mu$m spatial resolution of 21$''$ by convolving with an 
appropriate beam and regridding the resultant map to the same pixel scale. 
The final map is similar to the MIPS data, and the new peak position is indeed 
offset from that in the original PACS data by over 
10$''$. Because of the almost 4 times coarser spatial resolution of {\it Spitzer} 
at 70 $\mu$m, the true peak position had been smoothed out and merged with the 
surrounding PDR emission, and thus appeared shifted.\\
\begin{figure}
  \centering
  \includegraphics[width=4.5cm]{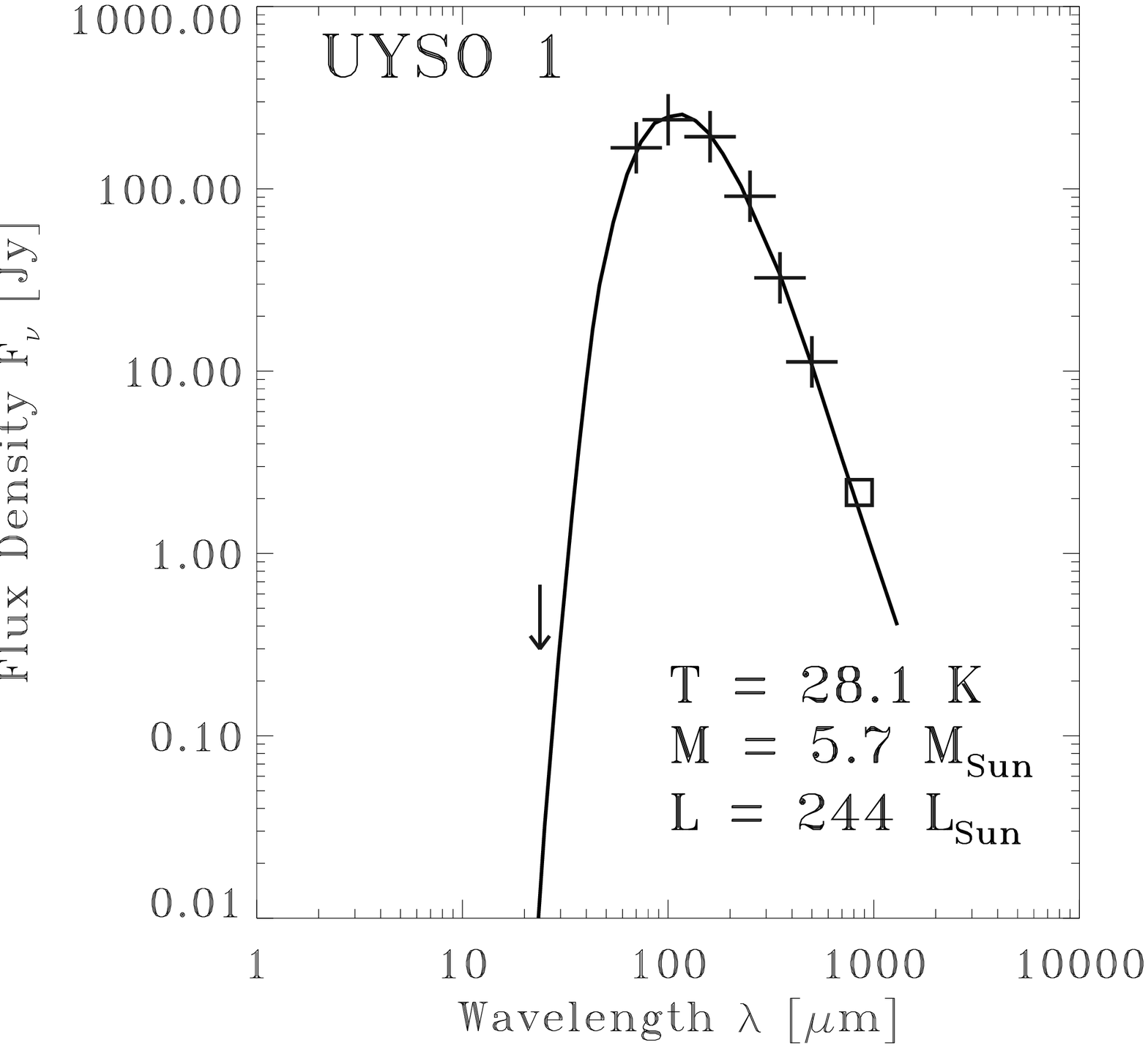}\includegraphics[width=4.5cm]{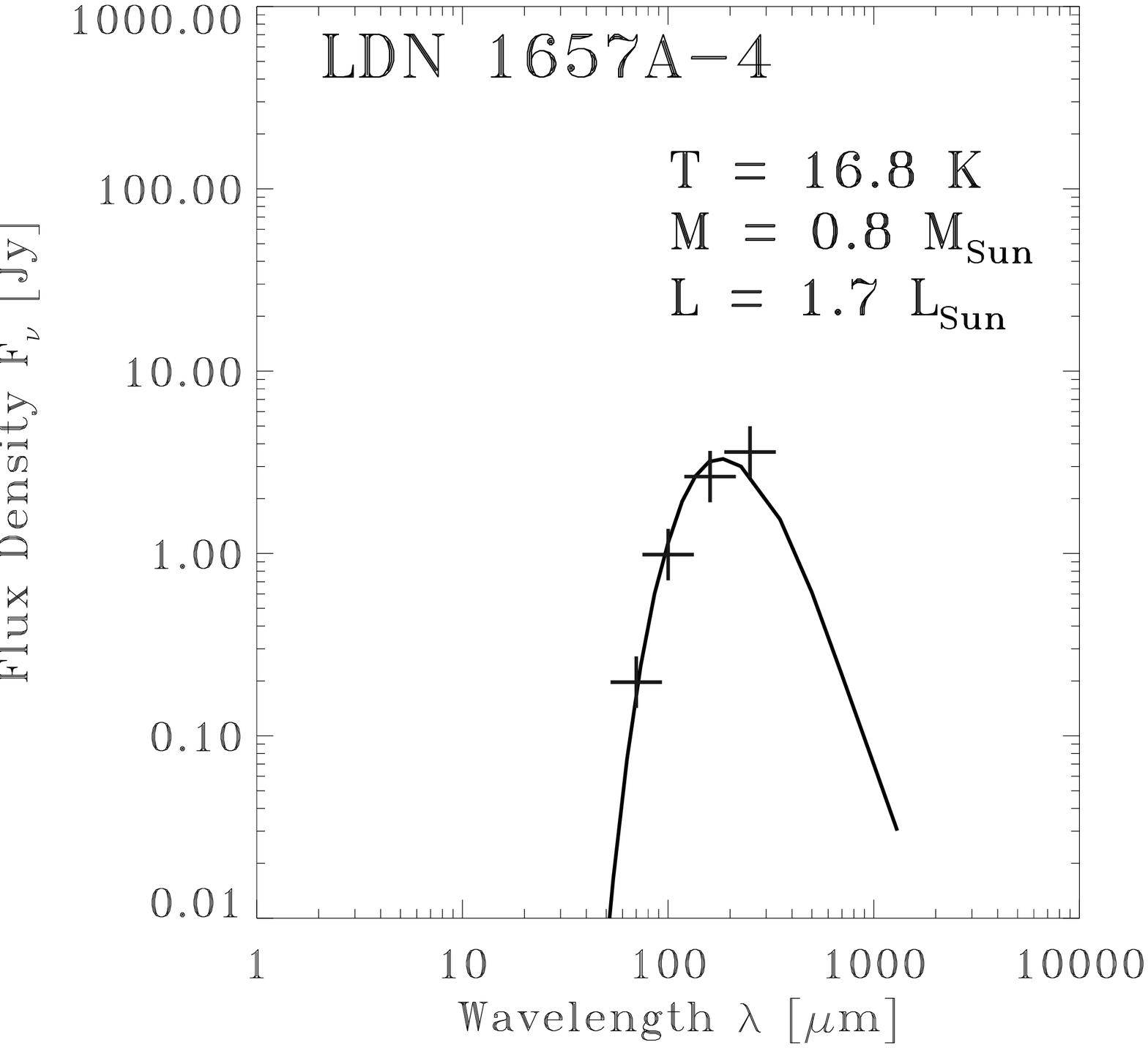}
  \caption{Single-temperature modified black-body SED fits {\bf left}
    for UYSO~1, and {\bf right} for LDN 1657A-4. Plotted on the same
    scale, this illustrates the different natures of the two regions.
    Both results were obtained by adopting the
    \citet{1994A&A...291..943O} opacities (their Table 1,
    $\kappa$-column 5). For T$\lesssim$ 20 K, colour correction terms
    for PACS fluxes become larger than the measurement uncertainties
    (mainly at 70 $\mu$m). They were applied to the LDN 1657A-4
    fluxes before the fitting.  {Plus signs: {\it Herschel} data, square: 
    SCUBA \citep{2004ApJ...602..843F}} }
  \label{Fig:UYSO-SED}
\end{figure}
   \begin{figure}
   \centering
   \includegraphics[bb=20 39 575 266,width=9cm,clip]{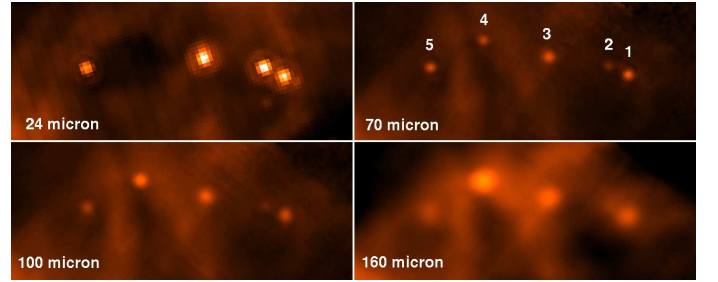}
      \caption{Compact sources in the LDN 1657A region
        (Table~\ref{Table:new-sources}), north-west of UYSO~1. The
        cutouts show a region of $220'' \ \times \ 90''$ size at 24
        $\mu$m (upper left), 70 $\mu$m (upper right), 100 $\mu$m
        (lower left), and 160 $\mu$m (lower right). Note especially
        the steep SED rise of source 4, spanning from a non-detection at
        24 $\mu$m to being the strongest source at 160 $\mu$m.  }
         \label{Fig:LDN1657A}
   \end{figure}
\indent
The comparison with the PACS 70 $\mu$m point-spread function (PSF) in
Fig.~\ref{Fig:70micron-centre} illustrates that UYSO~1 is not an
unresolved point source. The PACS PSF has a characteristic triangular
outer shape. 
However, UYSO~1 is marginally resolved at 70 $\mu$m; the slight elongation 
is along the connecting line between the two mm continuum sources. 
On the other hand, the stronger of the two outflows is also oriented in this
north-west/south-east direction \citep[cf.][ their
  Fig.~2]{2009A&A...493..547F}.  Therefore, both source
multiplicity and warm material along the outflow cone
can explain our findings. \\
\indent
The SED of UYSO~1 was estimated by fitting a Planck function, modified
by wavelength-dependent opacities, to the {\it Herschel} data. Our approach
was to use fluxes derived from uniformly sized apertures. 
We chose a circle with an area
equivalent to the SPIRE 500 $\mu$m beam area (1543$''^2$). We used
the opacities of coagulated grains with thin ice mantles
\citep{1994A&A...291..943O}. One temperature component was sufficient
to obtain a very good fit (see Fig.~\ref{Fig:UYSO-SED}). The derived 
temperature of $\sim$28.1 K was lower than the previous estimate of 40 K in
\citet{2009A&A...493..547F}, based on MIPS SED mode observations that
probably covered more of the adjacent warmer PDR material. Nevertheless, 
our photometry also includes a small contribution from the PDR, which
is heated from the outside, thus, the intrinsic luminosity is slightly 
overestimated 
\citep[in principal agreement with simulations performed by][]{2006ApJ...645..920S}. 
Thanks to the clarified astrometry, actual Herschel fluxes for $\lambda < 
450\,\mu$m could be included that exceed the {\it Spitzer} upper limits 
(which had been obtained after subtracting an extended-emission component) of 
\citet{2009A&A...493..547F}. Consequently, the derived luminosity (244 L$_\odot$) 
is higher than their estimate (50 L$_\odot$).
We note that the derived total mass of 5.7 M$_\odot$ strongly depends on 
the chosen dust model. We repeated the fitting with \citet{1984ApJ...285...89D} opacities, 
deriving instead a total mass of 26.4 M$_\odot$ (and $T$ = 26.1 K).

\subsection{New very cold source detections}\label{Sect:new-sources}

While UYSO~1 was the prime target of the {\it Herschel} observations, we also
discovered new sources $3\arcmin$ -- $4\arcmin$ to the north-west of
UYSO~1, within the neighbouring dark cloud LDN~1657A. From the SCUBA
850 $\mu$m map shown in \citet{2004ApJ...602..843F}, it was obvious
that this region is associated with submillimetre emission. The
PACS data resolve this region into a ``pearl necklace'' of five
compact sources (Table~\ref{Table:new-sources}).  Nothing is known
about these objects except the association of LDN 1657A-3 with faint
2MASS K band emission. There is no counterpart at 1.4 GHz in the NVSS
VLA survey.  Unfortunately, the region is at the very edge of
previous ROSAT X-ray observations \citep{2009A&A...506..711G} and thus
not properly covered. However, inspection of the entire dataset of 
MIPS 24 $\mu$m data\footnote{The region has not yet been observed with
  {\it Spitzer} at wavelengths $< 24 \ \mu$m. The compact sources are not
  detected in the MSX images.}, of which just the central part was 
published in \citet{2009A&A...493..547F}, reveals
that 4 of these 5 new sources are indeed detected
(cf.~Fig.~\ref{Fig:LDN1657A}).  At wavelengths $\le$160 $\mu$m, all
these sources are unresolved. Only at SPIRE wavelengths does a more
pronounced extended emission component become apparent
(Fig.~\ref{Fig:PACS-3colour}). Faint far-IR filaments seem to connect 
UYSO~1 and these northwestern sources. Hence, we assume that they are 
associated with the general star-formation complex.  
The distances between neighbouring LDN 1657A
sources are quite similar (39$''$\,--\,43$''$), only the two
western sources 1 and 2 being closer together (15$''$).  At a distance
of 1 kpc, this translates into around 0.2 pc (and 0.073 pc, respectively)
of projected distance. The distance 0.2 pc is roughly the Jeans length 
of a cold (16\,--\,18 K), medium-dense (1\,--\,2$\times10^4$ cm$^{-3}$) gas.
Thus, these sources most likely represent distinct
cores.  We report the PACS photometry in Table~\ref{Table:fluxes},
based on PSF photometry 
\citep[see also][]{A&ASpecialIssue-Henning}
using the Starfinder tool \citep{2000SPIE.4007..879D} and reference
PSFs provided by the PACS instrument team.  {Most of these sources
are detected at 24 $\mu$m, which is indicative of a second, warmer
dust component.}  \\
\indent
Source LDN 1657A-4 has the steepest
rising SED slope of all these sources 
(Figs.~\ref{Fig:UYSO-SED}, \ref{Fig:LDN1657A}). At 24 $\mu$m, the
point source was undetected;  we instead found a 24~$\mu$m
shadow at its location, caused by the high column density of the
obscuring material associated with the cold core.
We used the approach of \citet{2009A&A...499..149V} to
transform the 24 $\mu$m shadow contrast into a column density
$N$(H). The intensity ratio of the surrounding extended emission
to the darkest pixel in the 24 $\mu$m shadow was 1.51, after
subtraction of the estimated zodiacal foreground of 25.56 MJy/sr.  The
optical depth was then $\tau = \ln 1.51$, and the column
density was given by $N$(H) = $\tau / \sigma$.
We assumed the extinction cross-section $\sigma$ from the $R_{\rm V}$=5.5B
dust model\footnote{1.29$\times10^{-23}$ cm$^2$/H for MIPS~24 isophotal wavelength
  of 23.68 $\mu$m.} of \citet{2001ApJ...548..296W}, appropriate for
the denser shielded regions of molecular clouds.  We estimated a column
density of $3.2 \times 10^{22}$ cm$^{-2}$. This corresponds to 
$A_{\rm V}$=26 mag, using the \citet{1996Ap&SS.236..285R} calibration.
A more elaborate modelling of the fore- and background contributions
\citep[e.g.,][]{2007ApJ...665..466S} would probably increase the
calculated column density, but is beyond the scope of this letter.  
In addition, the standard (coarse) pixel scale  {of 2\farcs45} 
in the MIPS 24 $\mu$m data used as well as an increased level of
striping in the data limit the shadow contrast. Thus, the derived
24~$\mu$m shadow column density represents a lower limit.
Using the SPIRE 250 $\mu$m peak flux measured at the position of LDN 1657A-4
(7.23 Jy/ 18\farcs1 beam), the values of \citet{1994A&A...291..943O} 
opacities at this wavelength (18 cm$^2$/g), and the temperature 
derived from the
SED fit to the photometry data (17 K, see Fig.~\ref{Fig:UYSO-SED},
right), we also calculate a column density of $3.2 \times 10^{22}$
cm$^{-2}$. This value is certainly a lower limit, since the relatively
large SPIRE beam smears out the true column density
peak. Nevertheless, the very good agreement between the two independent
methods is compelling.  Finally, integration of the 24 $\mu$m 
extinction map over the SPIRE 250 $\mu$m beam area results in a total 
mass of 1.6 M$_\odot$, a higher value than that derived from SED fitting 
(0.8 M$_\odot$) but still comparable.

\section{Conclusions}

We have presented new {\it Herschel} PACS and SPIRE scan map data for the
star-forming complex containing the intermediate-mass core
UYSO~1. They show the dust emission structures associated with this
region in unprecedented detail. The high spatial resolution
of the PACS 70 $\mu$m map facilitates the differentiation of the true 
emission peak position from emission arising from the surrounding PDR material.  
The PACS data show that the 70 $\mu$m
emission is closely associated with the known location of the central
millimetre peak(s) in UYSO~1. This revises an
earlier finding of large offsets (based on {\it Spitzer}/MIPS 70 $\mu$m
data).  Hence, the measured {\it Herschel} photometry can be used with more
confidence than for previous data to estimate the SED of UYSO~1. Since the core was
not clearly detected at 24 $\mu$m (which otherwise would indicate a distinct
component of warmer dust), a one-temperature modified black-body
component with T = 28 K fits the data between 70\,--\,500 $\mu$m
very well. While clearly not in the hot-core/hot-corino regime,
this temperature is higher than the typical values of $<$17--20 K
found in other very young regions of star formation. The known outflow
activity can be one explanation. However, the influence of the
nearby PDR has also to be taken into account \citep{2009A&A...493..547F}.
The Herschel data reveal the serendipitous discovery of five compact
very red sources, situated to the north-west of UYSO~1, within the
neighbouring dark cloud LDN 1657A. These objects have lower
temperatures than UYSO~1. A particularly intriguing object is LDN 1657A-4,
which has a very red SED and exhibits a 24 $\mu$m shadow. We  
note that two different methods for the column density estimation provided
very consistent results. For both objects, the
\citet{1994A&A...291..943O} grain opacities seem more suitable than
the use of ISM grains.

\begin{acknowledgements}
PACS has been developed by an ins\/ti\-tu\-te consortium
led by MPE (Germany) and including UVIE (Austria); KU Leuven, CSL,
IMEC (Belgium); CEA, LAM (France); MPIA (Germany); INAF-IFSI/OAA/OAP/OAT, 
LENS, SISSA (Italy); IAC (Spain). It has been
supported by the funding agencies BMVIT (Austria), ESA-PRODEX (Belgium),
CEA/CNES (France), DLR (Germany), ASI/INAF (Italy), and CICYT/MCYT (Spain).
We thank the anonymous referee for a constructive report.
\end{acknowledgements}

\bibliographystyle{aa}
\bibliography{Linz}


\end{document}